\documentclass{aa}
\usepackage{amssymb}
\usepackage{amsmath}
\usepackage{natbib}
\bibpunct{(}{)}{;}{a}{}{,}
\usepackage{graphicx}

\begin{document}
\title{On the tidal evolution of the orbits of low-mass satellites around black holes}
\author{Andrej \v Cade\v z\inst{1}
\and Massimo Calvani\inst{2}
  \and Uro\v s Kosti\'c\inst{1}
 }
\offprints{A.\ \v Cade\v z, \email{andrej.cadez@fmf.uni-lj.si}}
\institute{Faculty of Mathematics and Physics, University of
Ljubljana, Jadranska 19, 1000 Ljubljana, Slovenia \and INAF -
Astronomical Observatory of Padova, Vicolo Osservatorio 5, 35122
Padova, Italy}
\date{Received 30 January 2008 / Accepted 21 May 2008}
\abstract {
Low-mass satellites, like asteroids and comets, are expected to be present around the black hole at the Galactic center. We consider small bodies orbiting a black hole, and we study the evolution of their orbits due to tidal interaction with the black hole. }
 {In this paper we investigate the consequences of the existence of plunging orbits when a black hole is present. We are interested in finding the conditions that exist when capture occurs.}
 {Earlier  analysis of the evolution of classical Keplerian orbits was extended to relativistic orbits around a Schwarzschild black hole.}
{The main difference between the Keplerian and black hole cases is in
the existence of plunging orbits. Orbital evolution, leading from
bound to plunging orbits, goes through a ``final" unstable circular
orbit. On this orbit, tidal energy is released on a characteristic
black hole timescale.} {This process may be relevant for explaining
how small, compact clumps of material can be brought onto plunging
orbits, where they may produce individual short duration accretion
events. The available energy and the characteristic timescale are
consistent with energy released  and the timescale typical of Galactic
flares.}
\titlerunning{On tidal evolution of orbits around black holes}
\authorrunning{\v Cade\v z et al.}
\maketitle
\section{Introduction}
Most galaxies, if not all, harbor massive black holes at their
center. Some of them manifest their presence in a violent way and
show up as active galactic nuclei; others, like the one in our
Galaxy, are quiescent. The mass of the black hole at the center of
the Milky Way has been estimated as $M_{\rm bh}=3.61\pm 0.32
\times 10^6 M_\odot$ from observations with SINFONI \citep{Eisenh05}
confined within a radius of 45 AU. The Galactic center is only
$\approx 8$ kpc from the Sun and therefore it is the nearest
laboratory where we can study the environment of super-massive black
holes in detail, see e.g.~\citet{2005PhR...419...65A},
\citet{2007Goldwurm}.

The standard assumption for AGNs is that gas accretes onto the black
hole forming an accretion disc and releasing energy. On the other
hand, some non-active galaxies exhibit short-timescale, low-intensity flaring activity
\citep{2003Natur.425..934G,2006JPhCS..54..420B,2006A&A...455....1E,2006A&A...460...15M},
which could be the result of individual accretion events of small
bodies onto the central black hole.

It seems reasonable to assume that stars at the Galactic center are
surrounded by planets and by other small orbiting bodies, like
asteroids and comets. Such low-mass satellites (LMS) may be stripped
off their parent stars by tidal interaction, while they approach the
black hole. In this way they contribute to a distribution of low-mass
objects that cluster the central black hole.
According to \citet{1995ApJ...455..342C}, the Edgeworth-Kuiper belt
of our Solar System may still contain as many as $2\times 10^8$
objects with radii $\lesssim$ 10~km. One might therefore expect that
LMS will be copiously present all the way down to the black hole.

In this paper we study the evolution of orbits of LMS caused by tidal
interaction with the black hole. We extend the analysis presented by
\citet{1980A&A....92..167H,1981A&A....99..126H,1982A&A...110...37H}
concerning the tidal evolution of
close binary systems in the weak friction model in the Newtonian
approximation. We show that orbital evolution of these objects
leads to capturing orbits by the black hole and argue that
tidal interaction can rapidly inject enough energy to explain the
energy source of Galactic flares.
\section{Low-mass satellites in the Galactic center}
\subsection{Energy-loss mechanisms}
The LMS are injected into the asteroid cloud with low
relative velocities with respect to the donor star, therefore their
average random velocity is roughly the same as that of stars. In
thermodynamic equilibrium, LMS would be expected to reach much higher
random velocities; however, it is not difficult to show that such
equilibrium cannot be reached, since the relaxation time $t_{\rm
a*}$ to exchange energy between stars and LMS is
\begin{equation}
  t_{\rm a*}={3\over 16}\sqrt{2\over \pi} {\sigma_{\rm a}^5\over c^4 n_* \sigma_*^2 r_{\rm g*}^2 \ln \Lambda} = \left ({\sigma_{\rm a} \over \sigma_*}\right )^5 T_{\rm h}\ ,
\end{equation}
where $\sigma_{\rm a}$ is the random velocity of LMS, $n_*$,
$\sigma_*$, and $r_{\rm g*}$ are the density, random velocity, and
gravitational radius of stars in the central cluster, $\Lambda $ is
the ratio of the maximum and minimum impact parameters, and $T_{\rm
h}$ is the non-resonant relaxation time from \citet[hereafter
HA06]{2006ApJ...645.1152H}. Since $T_{\rm h}$ is already $\sim 10^9\
{\rm yr}$, $\sigma_{\rm a}$ cannot increase much before the
relaxation time is longer than the Hubble time. This means that LMS do
not reach thermodynamic equilibrium random velocities, and the Bondi
radius of the LMS mass distribution is approximately the same as
that of the stellar distribution, i.e. $r_{\rm B}\sim 2\ {\rm pc}$
(HA06). However, the direction of angular momentum of such LMS
changes on a timescale shorter than $T_{\rm h}$, therefore the
distribution of these objects eventually becomes spherically
symmetric around the central black hole.

Just like stars, the LMS eventually accrete onto the central black
hole. To do so, they must lose orbital energy and/or
angular momentum. While stars eventually lose orbital energy  by
emitting gravitational waves, LMS are not massive enough for this
mechanism to play an important role. Considering the gravitational
radiation energy loss timescale \citep[Eq.~36.17b]{MTW}
\begin{equation}
t_{\mathrm{GW}}= \frac{5}{256}\frac{M_{\rm bh}}{m}\left(  {a\over r_{\mathrm{g}}}\right)^4
                                        {r_{\mathrm{g}}\over c}\gtrsim 10^{14}\ {\rm yr}\times \frac{10^{20}{\rm g}}{m}\ ,
\label{eq:tgw}
\end{equation}
where $M$ and $r_{\rm g}( = GM_{\rm bh}/c^2)$ are the mass and
gravitational radius of the black hole, $m$  the mass of the LMS,
and $a(>4r_{\rm g})$ the radius of its orbit. One finds that
satellites with mass less then $10^{23}{\rm g}$ can circle the
central black hole even at $a=4r_{\rm g}$ for the whole Hubble time.

However, hydrodynamic drag by circum-black-hole cloud might become
important in dissipating the energy of LMS. The timescale can be
estimated using similar arguments to \citet{2000ApJ...536..663N} to
obtain
\begin{align}
t_{\rm d} &= \frac{m}{\pi \rho_{\rm gas}R_{\rm eff}^3}\frac{R_{\rm eff}}{c}
                \left( \frac{a}{r_{\rm g}}\right)^{\frac{1}{2}}\nonumber\\
            &\gtrsim 3.55\ 10^{10}\ \mathrm{yr}
            \times\frac{10^3\ \mathrm{cm^{-3}}}{n_{\rm H}}
                            \left( \frac{m}{10^{20}\ \mathrm{g}}\right)^{\frac{1}{3}}\left( \frac{a}{4\ r_{\rm g}}\right)^{\frac{1}{2}}\ ,
\end{align}
where $R_{\rm eff}$ is the effective radius of LMS, and $\rho_{\rm gas}$
the density of the interstellar gas. In the second line we assume
that the satellite density is $5\ \mathrm{g/cm^3}$, $a=4\ r_{\rm g}$,
and $\rho_{\rm gas}$ is expressed in terms of hydrogen atom density.
The energy loss due to drag strongly depends on the density of the
gas and becomes comparable with nonresonant relaxation, if the mass
of LMS is below
\begin{equation}
m=2.24\ 10^{15}\ \mathrm{g}\times \left( \frac{n_{\rm H}}{10^3\ \mathrm{cm^3}}  \right)^3 \left({4\ r_{\rm g}\over a}\right )^{3\over 2}\ .
\end{equation}
For example, an object  circling at the Bondi radius in a gas with
density $n_{\rm H}\sim20\ \mathrm{cm^{-3}}$
\citep{2006ApJ...640..319X} must be less massive than $\sim 5\
\mathrm{g}$ for hydrodynamic drag to take over.
%
%
%
%
In the very vicinity of the black hole, the gas density may be higher
and also the factor $a/r_{\rm g}$ is much larger, yet, since there
is no evidence of an accretion disk \citep{2005PhR...419...65A}, it
does not seem plausible for these factors to increase enough for
hydrodynamic drag to play a very important role.

Since the evolution of orbital parameters is dominated by the
process with the shortest timescale, we estimate that nonresonant
relaxation occurring on a timescale of $\sim 10^9$ years (HA06) is
one of the most important mechanisms until the LMS approaches the
black hole to within a few Roche radii, where tidal interaction
takes over.
\subsection{The population of low-mass satellites}
It has been determined by \citet{2006ApJ...643.1011P} that the
density of stellar distribution increases as $1/r^2$ to within the
inner $1''$ of the Galactic center. At this distance, corresponding
to $2\times 10^5\ r_{\rm g}$, there is a sharp drop, interpreted by
HA06 as the radius where gravitational radiation extracts orbital
energy at a sufficient rate to clean the inner region. Since for
low masses gravitational radiation energy loss is negligible, we
expect that their density keeps increasing either as $n(r)\propto
r^{-7/4}$ (HA06) or as $n(r)\propto r^{-2}$
\citep{2006ApJ...643.1011P} deep down to the very vicinity of the
black hole, where tidal effects may extract orbital energy and
angular momentum. Note that there are no stars (very few) and no gas
(no steady emission lines) in the central region ($r<2\times 10^5\
r_{\rm g}$), thus  it is likely populated predominantly by low-mass
satellites. Hence, we assume that the dynamics of this central cloud
is dominated by the exchange of angular momentum by nonresonant
relaxation (HA06) and by tidal interaction between the object and
the central black hole. As the nonresonant relaxation timescale is
roughly position-independent and the tidal interaction grows
inversely proportional to a high power of the distance from the
black hole, the tidal interaction finally prevails in determining
the way in which accretion occurs.

One should expect a fair proportion of LMS to move on highly
eccentric, low-periastron orbits. Tidal forces do significant work on
such  satellites near periastron. This lowers their orbital energy and
starts the significant evolution of orbital parameters \citep{Gomboc05}.
In this context two classes of such satellites should be taken into
account: those that are gravity dominated (i.e. those whose
fundamental quadrupole frequency is $\nu_{\rm g} \approx 2 \sqrt{G
\rho / 3\pi}$) and those that are solid-state dominated (whose
fundamental quadrupole frequency is $\nu_{\rm
s}\approx\frac{1}{4}c_{\rm s}/R$). (Here $\rho$ is the density of the
body, $R$ its radius and $c_{\rm s}$ the speed of sound.) Taking
$c_{\rm s}\approx 5\ \rm{km/s}$ and $\rho\approx 5\ \rm{g\ cm^{-3}}$
as typical values, we find that the radius dividing the two classes
is close to the radius of the asteroid Ceres. Therefore all
gravity-dominated satellites should have about the same fundamental
quadrupole frequency, corresponding to the period of about 54
minutes. All smaller satellites should have shorter fundamental
periods. This means that gravity-dominated satellites start rapid
tidal evolution when their periastron reaches  $r_{\rm p} = (GM_{\rm
bh}/(2\pi \nu_{\rm g})^2)^{1/3}\approx 10 r_{\rm g}$.  Solid-state
dominated bodies may start  significant tidal evolution even closer
to the black hole.
\section{Tidal evolution of the orbits}
\label{sec:TidalEvolution} Significant tidal orbital evolution for
gravity-dominated solid bodies will start when their periastra reach
down to $\approx $9 $r_{\rm g}$. We show, however, that
solid-state dominated bodies are also strongly affected by tides and are
expected to be heated by them, so that at a certain stage, they are
expected to melt and also to become gravity-dominated.

The first stages of tidal evolution of the orbit can be investigated
using Hut's formalism for the spin-orbit evolution of the two-body
system
\citep{1980A&A....92..167H,1981A&A....99..126H,1982A&A...110...37H},
at least until relativistic regime is reached. The evolution is
governed by the parameter $\alpha$, which is the ratio of the
orbital and rotation angular momentum that the binary would have at
stable equilibrium, characterized by $a_0$, the radius of the stable
circular orbit and $\omega_0$, the orbital and spin frequency of
both bodies. The parameter $\alpha$ is determined by the (conserved)
value of the total angular momentum of the system by Eq.~(57) given
by \citet{1981A&A....99..126H}. We take  the spin of
both bodies into account, since it has been shown by \citet{2005PhRvD..72l4016F}
that ``the black hole absorbs angular momentum and energy at the
same rate as the moon's tidal field sends energy and angular
momentum into the hole's horizon". Thus we generalize Hut's equation
to
\begin{equation}\label{L1}
L = \mu(G{\cal M})^{1\over 2}((m_1r_{(1)g}^2R_1^2 + m_2r_{(2)g}^2R_2^2)/\mu)^{1\over 4}(\alpha^{1\over 4} +
\alpha^{-{3\over 4 }})\ ,
\end{equation}
where $m_1$, $R_1$, $r_{(1)g}$, and $m_2$, $R_2$, $r_{(2)g}$ are the
mass, radius, and gyration radius of the two bodies, ${\cal
M}=m_1+m_2$ and $\mu=m_1m_2/{\cal M}$.

The (conserved) angular momentum, calculated initially,  is
\begin{equation}\label{L2}
L = m_1r_{(1)g}^2R_1^2\omega_1 + m_2r_{(2)g}^2R_2^2\omega_2 + \mu(G{\cal M}r_0)^{1\over 2}\ ,
\end{equation}
where $\omega_1$ and $\omega_2$ are the two initial spin angular
velocities, $h=\mu(G{\cal M}r_0)^{1/2}$ is the initial orbital
angular momentum with  $r_0=a(1-e^2)$, and $a$ and $e$ are the
semi-major axis and eccentricity of the initial orbit, respectively.
Let $m_1\gg m_2$  and
$R_1\hspace{2pt} r_{(1)g}\gg R_2\hspace{2pt} r_{(2)g}$
and define
\begin{align}
Z &= (m_2r_{(2)g}^2R_2^2)/(m_1r_{(1)g}^2R_1^2) \\
Y &= h/(m_1r_{(1)g}^2R_1^2\omega_1)\ .
\end{align}
Equating Eqs.~(\ref{L1}) and (\ref{L2}), we obtain
\begin{align}
\nonumber
(\mu/m_1)^{3\over 4} &(G{\cal M}/R_1^3)^{1\over 2} (1 + Z)^{1\over 4} (\alpha^{1\over 4} + \alpha^{-{3\over 4}})  \\
&=\omega_1(1 + Z\omega_2/\omega_1 + Y)\ .
\end{align}
Note that $\Omega_{\rm d} = (G{\cal M}/R_1^3)^{1/2}$ is the angular
frequency at which $m_2$ would orbit $m_1$ at its gyration radius.
If one assumes that $m_1=M_{\rm bh}$ and $m_2=m$, then
$\omega_1/\Omega_{\rm d} = a_{\rm Kerr}/r_{\rm g}c$ where $a_{\rm
Kerr}$ is the Kerr angular momentum parameter. Since  $M_{\rm bh}\gg
m$, we may replace ${\cal M}=M_{\rm bh}$,  $\mu =m$ and write
$R_2=R$. Finally, from the above equations we obtain
\begin{equation}
(\alpha^{1\over 4} + \alpha^{-{3\over 4}}) ={ a_{\rm Kerr}\over r_{\rm g}c} \left(M_{\rm bh}\over m\right)^{3\over 4}(1 + Z\omega_2/\omega_1 +
Y) (1 + Z)^{-{1\over 4}}\ .
\end{equation}
From this expression it is clear that the angular momentum of the
black hole determines $\alpha$; moreover, since $a_{\rm Kerr}/r_{\rm
g}c$ is not expected to be exceedingly small and $Z$ and $Y$ are
very small, the equation for $\alpha$ is essentially
\begin{equation}\label{alpha}
(\alpha^{1\over 4} + \alpha^{-{3\over 4}}) = { a_{\rm Kerr}\over r_{\rm g}c} \left(M_{\rm bh}\over m\right)^{3\over 4}\ .
\end{equation}
As the righthand side is very large, it follows that $\alpha$ is
either a very small or a very large number. Since, by definition,
$\alpha = h_0[(m_1 r_{(1)g}^2 R_1^2+m_2 r_{(2)g}^2
R_2^2)\omega_0]^{-1}~\approx ~
{1\over 4}(m/M_{\rm bh})(a_0/r_{\rm g})^2$
it is clear that we are not interested in very high,
but in very low values of $\alpha$ \footnote {The
equilibrium orbit is circular with radius $a_0$, so that $h_0=m
a_0^2 \omega_0$ and, since it is co-rotating with the black hole,
$\omega_0=\omega_{\rm Kerr}$. With respect to the moment of inertia
of the black hole $m_1 r_{(1)g}^2 R_1^2=J_{\rm bh}/\omega_{\rm
Kerr}$, the moment of inertia of the asteroid can be neglected.
According to \citet{Ash}, $J_{\rm bh}/\omega_{\rm Kerr}=M_{\rm
bh}R_{\rm h}^2$, where $R_{\rm h}$ is the horizon radius of the
black hole, which becomes $2r_{\rm g}$ in the limit of small $a_{\rm
Kerr}$.}. Hut was not interested in such orbits. We note that, if
$r_{\rm p}(\varepsilon \vert \alpha^\prime)$ is the solution to the
equation (27) in \citet{1982A&A...110...37H} for $\alpha^\prime>3$
and if $\gamma r_{\rm p}(\varepsilon \vert \alpha)$ is the solution of
the same equation for $\alpha<3$ such that $\alpha^\prime $ and
$\alpha$ are the two solutions of Eq.~(57) in
\citet{1981A&A....99..126H}, then a scaling factor $\gamma $ exists,
such that both are solutions of the same equation\footnote{ The
exchange $\alpha \rightarrow \alpha^\prime$ changes the unstable
equilibrium state $\tilde r_-$ to the attractor $\tilde r_{\rm p}=1$
and vice versa with respect to the tracks shown in
\citet{1982A&A...110...37H}.}. Therefore both $\alpha$'s solving
Eq.~(\ref{alpha}) produce the same solutions.

\begin{figure}
\includegraphics[width=4.48cm]{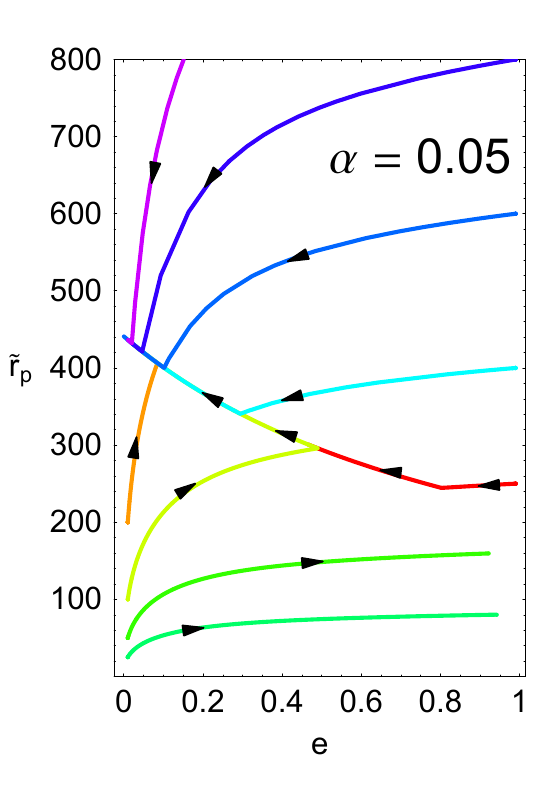}
\includegraphics[width=4.32cm]{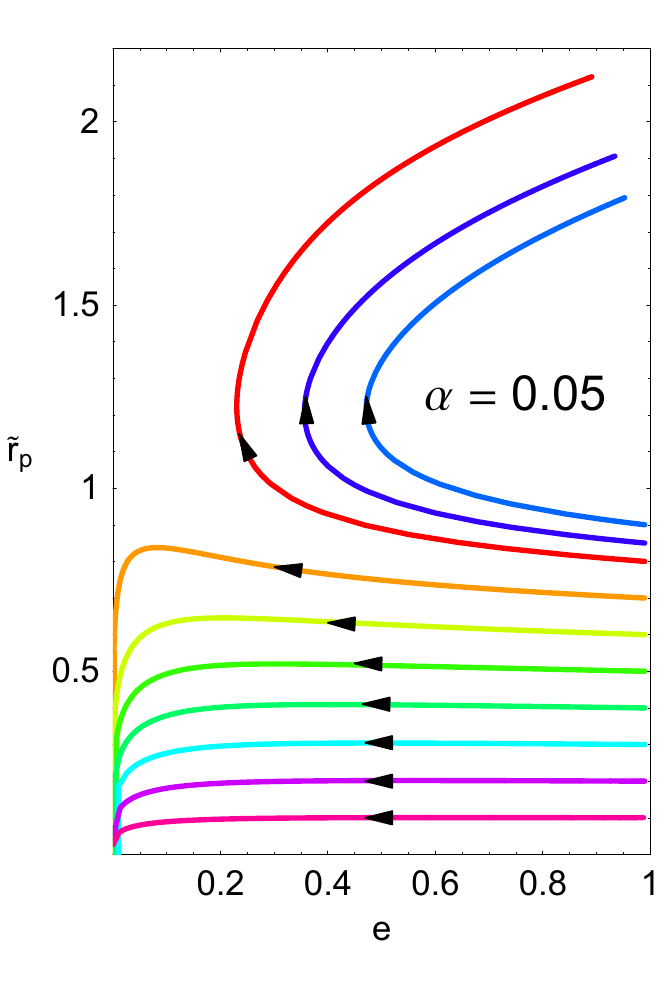}

\caption{Orbital evolution tracks in the eccentricity (e)-periastron
distance($r_{\rm p}$) plane. Different types of evolution are
displayed. Arrows indicate the direction of time. The left panel
shows tidal evolution toward stable corotating circular orbits at
$r_0^\prime~=450r_{\rm g}$.  The right panel is an enlargement of the
left bottom part. It shows the evolution of very low orbital angular
momentum orbits. (See text for details.) In this example, intended to show the topology of
small $\alpha$ tidal evolution, the value of $\alpha$ is arbitrarily
taken to be 0.05. Different curves correspond to initial $\tilde
r_{\rm p}=25, 50, 100, 200, 250,  400, 600, 800, 1000 $ (left panel)
and $\tilde r_{\rm p}=0.1, 0.2, 0.3, 0.4, 0.5, 0.6, 0.7, 0.8, 0.85,
0.9$ (right panel).
%
} \label{HutPE}
\end{figure}

\begin{figure}
\centering
\vspace{0.19cm}
\includegraphics[width=6.6cm]{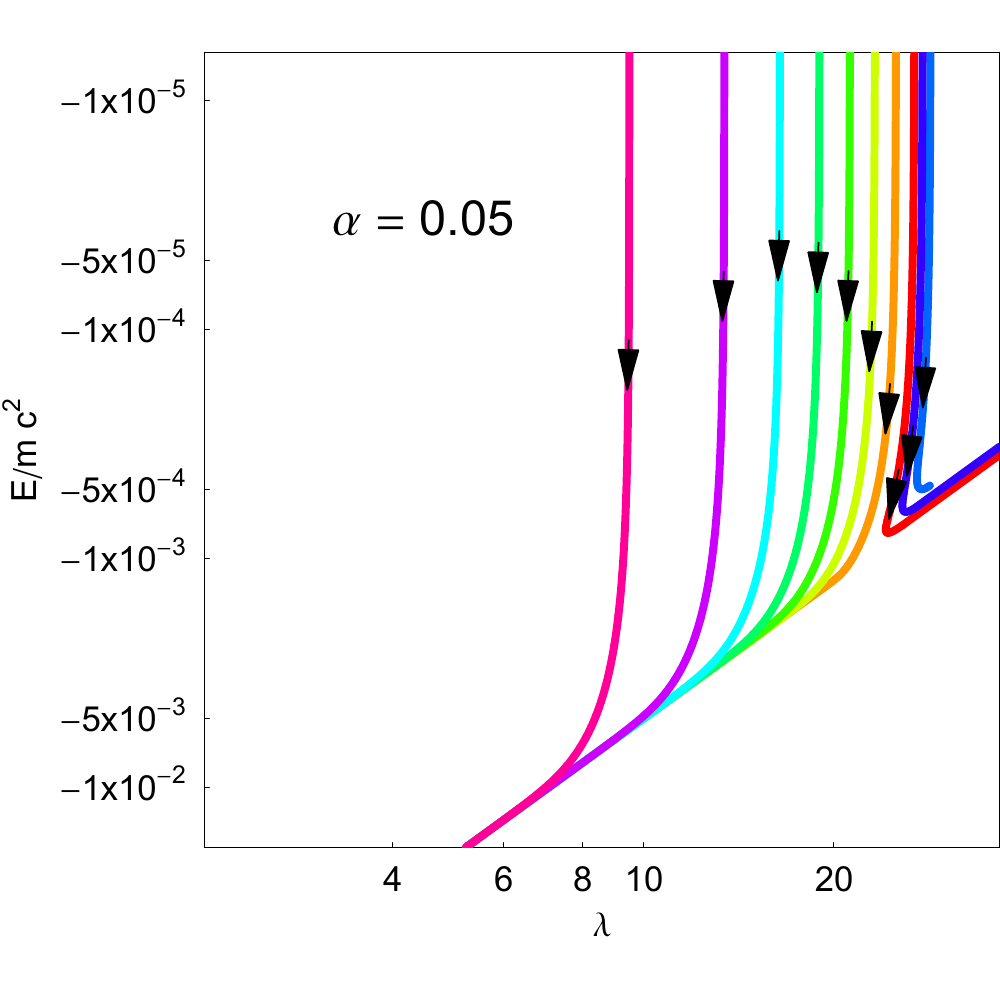}
\caption{Angular
momentum ($\lambda={h\over 2 m r_{\rm g} c}$) - orbital energy ($E$)
during the tidal evolution of the orbit for the evolution tracks
shown in the right panel of Fig. \ref{HutPE}.} \label{HutEL}
\end{figure}

The equation for the flow of trajectories in the ($\tilde r_{\rm p},
e)$ plane  for very low values of $\alpha$ is the same as that for very
high values of $\alpha $ and can be written using Eq.~(27) in
\citet{1982A&A...110...37H}
\begin{equation}
\frac{d\tilde r_{\rm p}}{d\varepsilon} = \frac{\tilde r_{\rm p}(-49 + 81\varepsilon -70\varepsilon^2 +
22\varepsilon^3 -17\varepsilon^4 + \varepsilon^5)}{11(-2 + \varepsilon)(-1 +
\varepsilon)(21 - 28\varepsilon + 18\varepsilon^2 -4\varepsilon^3 +\varepsilon^4)}\ .
\end{equation}
Here $\tilde r_{\rm p}=r_{\rm p}/a_0$ is the normalized minimum
separation of the two objects at periastron, $\tilde r_{\rm p} =
a(1-e)/a_0$. Noting that $\varepsilon=1-e$, where $e$ is the
eccentricity of the orbit, the above may be integrated to
\begin{align}
\nonumber
\tilde r_{\rm p} = &(21 - 28(1- e ) +18(1 - e)^2 -4 (1 - e)^3 + (1-e)^4)^{2\over 11}\\
&\times e^{4\over 11}\left (1 +
e\right )^{-1}\tilde r_{\rm n} \ ,
\label{rpevol}
\end{align}
where $\tilde r_{\rm n}$ is the integration constant. The meaning of
these limiting solutions for very small $\alpha$ is illustrated in
Fig.~\ref{HutPE} by the numerical solution of Eq.~(27) in
\citet{1982A&A...110...37H}.
In this example, where the value of $\alpha$ was chosen equal to
$0.05$ ($\alpha^\prime=9720$)\footnote{This is an unreasonably large
number with respect to our problem, but sufficiently small and
convenient to show the topology of the solution space.}, the
equilibrium radius $a_0^\prime$ is at $\tilde r_{\rm p}\approx 450$
and only orbits with $\tilde r_{\rm p}<0.8$ shrink as they
circularize.
In Fig.~\ref{HutPE}, the left panel
shows tidal evolution toward stable corotating circular orbits at
$r_0^\prime~=450r_{\rm g}$. These orbits are initially quite
eccentric and have low spin (the upper five curves). The remaining
orbits start as circular but with large spin. If the initial spin
energy is high enough, it can be transferred to orbit, first
elongating it and then circularizing again after joining the
circularization track (two upper curves starting to the left). If the
initial spin energy is not high enough, the circularization track
cannot be reached and the orbit keeps elongating at a slower and slower
pace (lower two curves). The right panel is an enlargement of the
left bottom part. It shows the evolution of very low orbital angular
momentum orbits. Very high initial spin angular momentum and energy
can be transferred to orbital momentum and energy, leading to higher
periastron orbits (the upper three curves). In the case of small
initial spin, the orbital evolution starts with orbital energy
dissipation by tidal interaction and little angular momentum
transfer, leading to less eccentric orbits with shorter and shorter
orbital periods. As a result, the orbital angular momentum
eventually transfers to spin as the object is forced into faster and
faster corotation.
As $\alpha^\prime \rightarrow \infty$, the
corresponding $a_0^\prime$ also tends  to infinity and only the
shrinking orbits remain as candidates to bring the body down to the
horizon of the black hole. With the solution for $r_{\rm p}$ and
noting that
\begin{align}
h &= m(G M_{\rm bh} r_0)^{1\over 2} = m r_{\rm g} c ((1+e) r_{\rm p}/r_{\rm g} )^{1\over 2}\\
 E &= - G M_{\rm bh}m (1 - e)/2r_{\rm p}\ ,
\end{align}
one can calculate  tracks of orbital evolution in the angular
momentum - energy plane (see Fig. \ref{HutEL}).
Using these tracks, one can calculate the evolution of orbital
radial turning points.

\begin{figure}
\includegraphics[width=4.45cm]{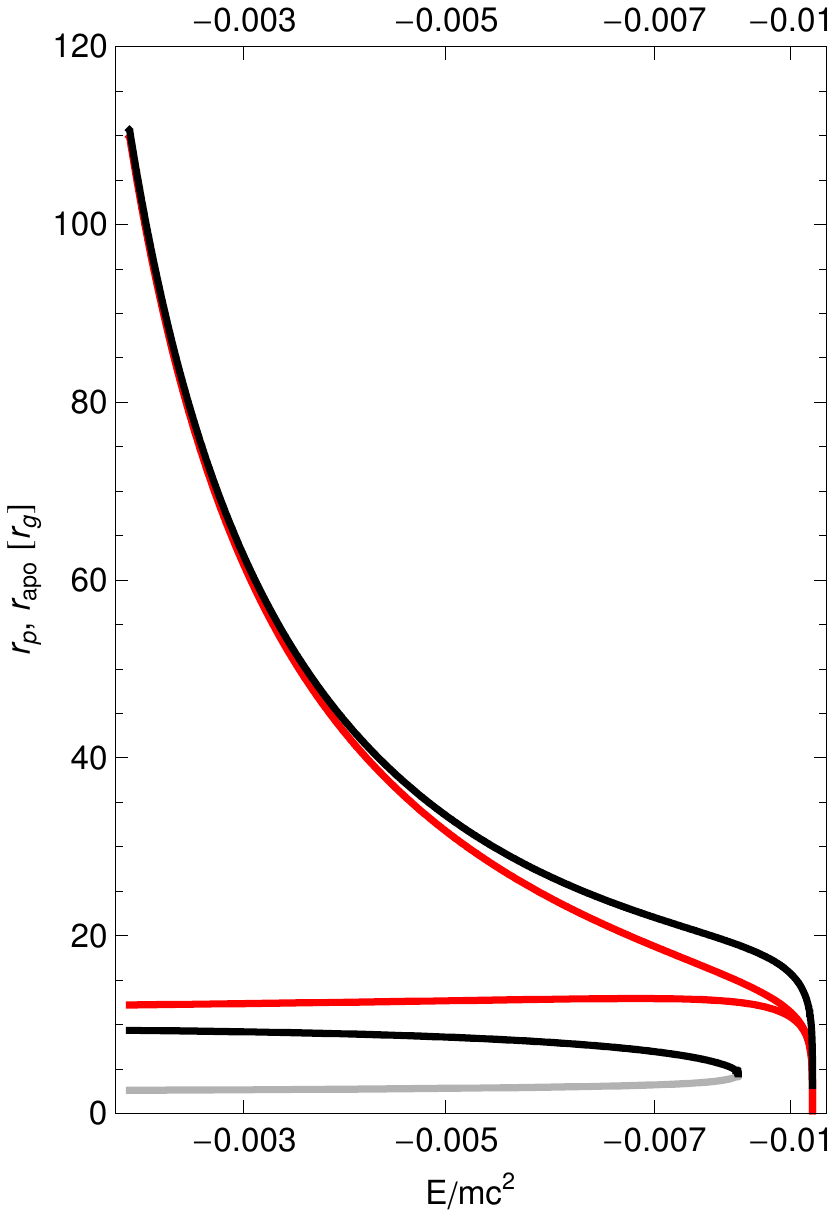}
\includegraphics[width=4.35cm]{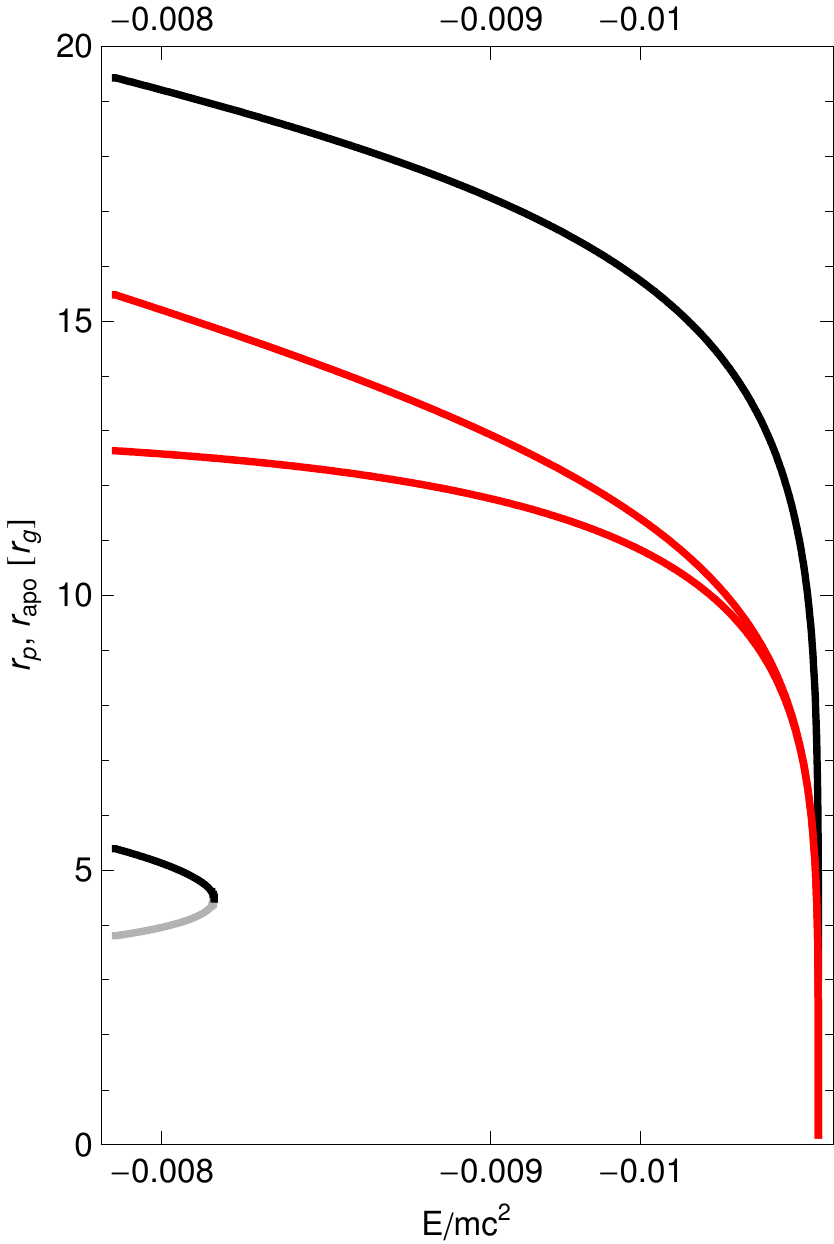}
\caption{The evolution of radial turning points in the Kepler (red) and
Schwarzschild effective potentials (black) for the case of a highly eccentric orbit starting with periastron at $9.5 r_{\rm g}$ and apoastron at $110 r_{\rm g}$. The right figure is an enlargement of the final evolution. The gray line indicates the third root of the effective potential equation \citep{MTW}.}
\label{PolyZero}
\end{figure}

We do this using both the Keplerian and the black hole effective
potentials. For simplicity we only consider the Schwarzschild metric
(Fig.~\ref{PolyZero}).
The important qualitative difference between the two cases occurs
for low values of orbital energy and angular momentum. In the
Keplerian case the effective potential guarantees two turning points
for all bound orbits, since it has one minimum and no maxima, while
in the Schwarzschild case the effective potential has a minimum
($V_{\rm min}=V(r_{\rm min})$) and also a maximum ($V_{\rm max}$) at
small $r_{\rm max}<r_{\rm min}$ \citep{MTW}. Therefore, the inner
turning point disappears in a parametric family of orbits, when the
effective potential maximum becomes less than the orbital energy.
For the parametric family of orbits shown in Fig.~\ref{HutEL}, this
happens when $r_{\rm max}$ is just slightly larger than $4r_{\rm
g}$. In Fig.~\ref{PolyZero} we then show that, from the point of view of
turning points, Schwarzschild orbits do not necessarily circularize,
but the inner turning point disappears when the outer one may still
be at $\approx 20r_{\rm g}$. The relativistic theory of tidal orbital
evolution may produce somewhat different tracks in the $\lambda -E$
plane, yet we believe it is reasonable to expect the results not to
change qualitatively, because the exchange between orbital and
internal angular momentum and energy occurs very locally in the
small volume of the tidally distorted body. In any case, when the
total energy of the body closely approaches the maximum value of the
effective potential ($V_{\rm max}$), the body experiences so shallow
an effective potential that it does not return to the apoastron, but
winds about $r_{\rm max}$.
\section{Time scales for tidal evolution}
\label{sec:TimeScales}

To estimate the timescale for tidal evolution of the orbits,
consider  the energy loss per orbit as given by Eq.~(A10) of
\citet{1981A&A....99..126H}\footnote{For the sake of simplicity we
neglect the contribution of work done by the tangential component of
tidal force, which depends on the difference between the orbital and
spin period.}. It can be written as
\begin{align}
\nonumber
\Delta E =& -\left( {{G m M_{\rm bh}R^2}\over{r_{\rm p}^3}} \right)
 {{9\pi}\over{(1+e)^{15\over 2}}}
  k e^2\\
  &\times (1+{15\over 4}e^2+{15\over 8}e^4+{5\over 64}e^6)
 {\sqrt{{G M_{\rm bh}}\over{r_{\rm p}^3}}} \tau
 \left(
{{r_{\rm R}}\over{r_{\rm p}}}
 \right)^3\ ,
\end{align}
where $r_{\rm R}=(M_{\rm bh}/m)^{1/3}R$ is the Roche radius. The
nonresonant energy timescale is then
\begin{equation}
t_{\rm EN} = - t_{\rm orb}{{GM_{\rm bh}m}\over{2a}}/\Delta E
\label{NRTScale}
\end{equation}
which becomes
\begin{equation}
t_{\rm EN}={64(1+e)^{15\over 2}r_{\rm p}^8\over{9 c^2\sqrt{1-e}\hspace{2pt} e^2(64+240\hspace{2pt}e^2+120\hspace{2pt}e^4+5\hspace{2pt}e^6)k R^2 r_{\rm g} r_{\rm R}^3\tau}}\ ,
\label{NonResCas}
\end{equation}
where $t_{\rm orb}$ is the orbital period, $k = 0.75$ the apsidal
motion constant of the primary (considered as an incompressible
fluid), and $\tau$ a constant small time lag
\citep{1981A&A....99..126H}.

\begin{figure}
\centering
\resizebox{\hsize}{!}{\includegraphics{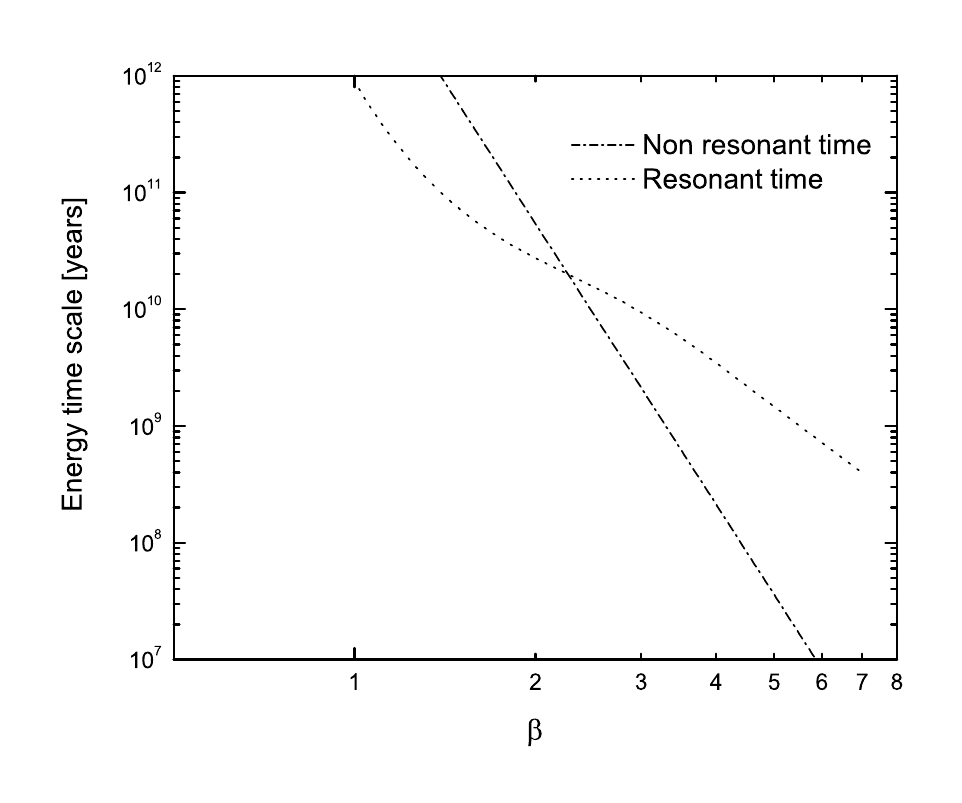}}
\caption{Resonant and non-resonant timescales as a function of $\beta$ for $\rho=\rho_0$, $R=10$km, and $Q=100$ (see Eqs.\ \ref{RTS},\ref{NRTS}).}
\label{ResNonRes}
\end{figure}

Remember, however, that Hut only considers weak tides acting on the
deformed body $m$. Such tides are quite off resonance, and this would
not be the case in the proximity of the black hole, where tides
would certainly be resonant if the body liquefies. Therefore a
better estimate is as follows. For resonant tides the body
liquefies. \citet{Gomboc05} obtain the following equation for tidal
energy per periastron passage, valid in the limit of high
eccentricity,
\begin{equation}
\Delta E = \left({{G M_{\rm bh} mR^2}\over{r_{\rm p}^3}}\right)\varepsilon^2(\beta) \ ,
\end{equation}
with the resonant timescale for tidal evolution,
\begin{equation}
t_{\rm ER} = t_{\rm orb}{{(1-e)r_{\rm p}^2}\over{2R^2\varepsilon^2(\beta)}}\ ,
\label{ResCas}
\end{equation}
and $\varepsilon^2(\beta)$ is a function that peaks at resonance,
i.e. at $\beta \approx 1$. Here $\beta =r_{\rm R}/r_{\rm p}$ is the
Roche penetration parameter. For easier comparison we rewrite
Eqs.\ \ref{NonResCas} and \ref{ResCas} into a simpler form
\begin{align}
t_{\rm EN} &= 1.4\ 10^{11}\ \mathrm{yr}\times\frac{Q f_{\rm EN}(e)}{\beta^8}\left(\frac{10\ \mathrm{km}}{R} \right)^2\left(\frac{\rho_0}{\rho}\right)^{7\over 6}
\label{RTS}
\\
t_{\rm ER} &= 2.9\ 10^{10}\ \mathrm{yr}\times \frac{f_{\rm ER}(e)}{\varepsilon^2(\beta)\beta^{7\over 2}}\left(\frac{10\ \mathrm{km}}{R} \right)^2\left(\frac{\rho_0}{\rho}\right)^{7\over 6}\ ,
\label{NRTS}
\end{align}
where $Q$ is the resonant damping factor, $\tau=(\omega_0 Q)^{-1}$,
$\omega_0=\sqrt{GM_{\rm bh}/r_{\rm R}^3}$, and $\rho_0=1\mathrm{g\
cm^{-3}}$. The function $f_{\rm EN}(e)$ is
\begin{equation}
f_{\rm EN}(e) =\frac{16(1+e)^{15\over 2}}{25\sqrt{1-e}\hspace{2pt} e^2(64+240\hspace{2pt}e^2+120\hspace{2pt}e^4+5\hspace{2pt}e^6)},
\end{equation}
and $f_{\rm ER}(e)$ (valid only for highly eccentric orbits) is
\begin{equation}
f_{\rm ER}(e) = \frac{1}{\sqrt{1-e}}\ .
\end{equation}
Both functions are of order unity on the interval $[0.1,0.9]$. In
Fig.~\ref{ResNonRes} we plot  $t_{\rm EN}$ and $t_{\rm ER}$ as a
function of $\beta$ for $R = 10$ km, $Q=100$ and $\rho=\rho_0$,
assuming that the body is liquefied, i.e. gravity-dominated.
For example, if an $R=30\ \mathrm{km}$ object with $\rho=5\
\mathrm{g\ cm^{-3}}$  is on a $\beta= 2$ orbit, then $t_{\rm
ER}=3\times 10^{10}\ \mathrm{yr} \times(10/30)^2 (1/5)^{7/6}=
5.1\times 10^8\ \mathrm{yr}$;  for this case $t_{\rm EN}$  is
approximately twice as long.
\section{The fate of a small body}
Consider now a cold solid object that is scattered on an orbit with
a periastron of a few 10 $r_{\rm g}$. Tidal energy loss and the
corresponding energy timescale can be accurately calculated using
Hut's formalism, since tides on such an object would be well below
resonance. However, one must replace the apsidal motion constant $k$
with a lower value and $\omega_0$ with the higher angular
frequency of the quadrupole mode of the solid body. Thus, to obtain
the energy time for such a solid body, the timescale in
Eq.~(\ref{NonResCas}) should be multiplied by the ratio
$\left(\nu_{\rm s}/\nu_{\rm g}\right)^3$, to obtain
\begin{equation}
t_{\rm ES} = 5.7\ 10^{19}\ \mathrm{yr}\times\frac{Q f_{\rm EN}(e)}{\beta^8}\left(\frac{10\ \mathrm{km}}{R} \right)^{5}\left(\frac{\rho_0}{\rho}\right)^{8\over 3}\ .
\end{equation}
The average heating power is
\begin{align}
P_{\rm heat} &= \vert \Delta E\vert/t_{\rm orb}\\ \nonumber &=~{1\over 2}GM_{\rm bh}m t_{\rm ES}^{-1}a^{-1}\ ,
\end{align}
where the last equality follows from Eq.~(\ref{NRTScale}). The
typical thermal diffusion timescale is $\tau_{\rm
solid}=R^2\pi^{-2}D^{-1}$, where $D$ is the thermal diffusion
coefficient, which is on the order of $10^{-6}\ {\rm m}^2{\rm
s}^{-1}$ for most rocks, so that $\tau_{\rm solid}\approx 3\times
10^5$ years for a $10$ km object. It is a simple exercise to
calculate the central temperature that such an object would reach in
equilibrium when the heating power is thermally diffused to the
surface, which is black-body radiating. Neglecting the surface
temperature, one obtains $T_{\rm c}=P_{\rm heat}(8\pi R \lambda_{\rm
c})^{-1}$, where the heat conductivity $\lambda_c\approx
1-2\mathrm{W m^{-1}K^{-1}}$ is typical of most rocks. Thus rocks
could reach central temperatures on the order of
\begin{equation}
T_{\rm c} = 0.02 K\  {1-e\over Q f_{\rm EN}(e)}\left({R\over 10\ {\rm km}}\right)^{7}\left({\rho\over \rho_0}\right)^4 \beta^9 \ .
\end{equation}
Assuming that natural values for $Q$ are a few hundred up to a few
thousand, the same as for normal modes of Earth, and noting the
extremely steep dependence of $T_{\rm c}$ on $R$, $\beta$, and
$\rho$, one expects a few 10 kilometer rocks to melt ($T_{\rm c}
\gtrsim 3000K$) at their centers a few hundred thousand years after
entering a $\beta \gtrsim 2$ orbit. Melting weakens the body,
increases its coupling to tides, and intensifies heating. Therefore,
the core melts farther out, but does not essentially increase its
temperature, because the convection that sets in is extremely efficient
in transporting  heat from the center to the remaining solid
envelope. This runaway process stops when further melting can no
longer weaken the body, i.e. when the body is no longer solid state
but gravity-dominated. If melting had started at $\beta \gtrsim 1$,
then the body finds itself at tidal resonance, and the pertinent timescale
shortens considerably and becomes $T_{\rm ER}$
(Eq.~\ref{NRTS}). The surface temperature of the body is now
determined by the balance between tidal heating and black body
cooling, so it can be expressed as
\begin{equation}
T_{\rm bb}=610\ K \left ({(1-e)\varepsilon ^2(\beta)\over f_{\rm ER}(e)}\right )^{1\over 4} \left( {R\over 10\ {\rm km}}\right)^{3\over 4}\left({\rho \over \rho_0}\right)^{5\over 8}\beta^{9\over 8}\ .
\end{equation}
Because of the $1/4$ power in Stefan's law, the black body surface
temperature is a slowly varying function of $R $, $\beta$, and
$\rho$. Roughly speaking, it is expected to be just below the
boiling temperature of rocky material (a few thousand Kelvin) for
melted rocks at a tidal resonance whose size is between about 10 and
100 kilometers. Much smaller rocks would not melt at all, and much
larger rocks, moons, and planets would melt farther away from the
black hole and would evaporate or split into smaller pieces or drops
when reaching the Roche radius. Evidence of a similar tidal melting
may also be found in our Solar System \citep{1988Icar...76..295D}.

Heating power increases  slowly with slowly increasing $\beta$, and
the melted body undulates as it moves between periastron and
apoastron on an  eccentric enough orbit that keeps it outside the Roche radius most of
the time.

The above considerations apply until angular momentum transfer
becomes important. Angular momentum can be transferred from orbital
to spin of the black hole and from orbital to spin of the body. The
angular momentum transfer to a Schwarzschild black hole  has been
studied by \citet{2005PhRvD..72l4016F} for the case of tidal
coupling of a black hole and a circularly orbiting moon. They find
that tidal interaction gives rise to orbital angular momentum loss
rate (see their Eq.~27):
\begin{equation}
\dot h={32 \over 5} {m\over M_{\rm bh}}{r_{\rm g}^7\over r^8} h c\ .
\label{Lovel}
\end{equation}
In our case $m/M_{\rm bh}$ is so small that the timescale resulting
from Eq.~(\ref{Lovel}) is longer than the Hubble time, so this
effect is negligible. On the other hand, the angular momentum
transfer to the spin of the body can work. We are now considering an
already melted electrically conducting object that is constantly
elongated by the mounting tidal force. Therefore, its moment of
inertia is increasing, while the ever longer object is spinning with
the orbital angular velocity. This process only has meaning if the
body remains a whole and sufficiently rigid, i.e. if there is a
mechanism that distributes angular momentum to all parts of the
body. In our case sufficient rigidity can be provided by a frozen-in
magnetic field. The necessary field must be strong enough to insure
that Alfv\' en waves can travel the length ($l$) of the body before
the frozen-in magnetic field decays, i.e. $v_{\rm A} \tau_{\rm
m}>l$. Here $v_{\rm A}\sim B/\sqrt{\mu_0 \rho}$ is the Alfv\' en
velocity and $\tau_{\rm m} = \sigma \mu_{0} l_t^2$ is the magnetic
field decay time, where $l_{\rm t}$ is the transverse dimension of
the body. Taking a typical value $\sigma\sim 4 \times 10^6\
\Omega^{-1}m^{-1}$ for the conductivity, one finds that a field of
$\sim100\ \mu \mathrm{G}$ is enough to give rigidity to a $100\
\mathrm{km}$ object. Such and much stronger fields have been found
in asteroids \citep{1993Sci...261..331K}, so that it is not
unreasonable to assume that the LMS in the vicinity of the Galactic
black hole also possess it.

Slow orbital angular momentum decay describing the lower part of
tracks in Fig.\ref{HutEL} has no dramatic consequences until fully
relativistic regime is reached, i.e. until the effective potential
maximum starts rapidly dropping with loss of angular momentum. This
happens when the inner turning point approaches $r_{\rm max}\approx
4r_{\rm g}$. The shallow potential no longer returns the body from
periastron to apoastron, but the orbit just keeps winding about
$r_{\rm max}$ while the constant large tidal tensor, now experienced
by the body, keeps squeezing and pulling it apart into a long thread
along the orbit, exponentially increasing its internal energy.

We note that the above analysis is an approximation somewhat limited
by neglecting dynamical effects and relativistic aspects of tides.
As pointed out in Sect.~\ref{sec:TidalEvolution}, in the relativistic
(Schwarzschild) regime, the orbit has no inner turning point when
the effective potential maximum becomes less than the orbital
energy, leading therefore to capture orbits with orbital energy that
may be considerably higher than the orbital energy of the marginally
stable orbit. This energy difference is available to do tidal work
on the body before entering the final turn down the horizon of the
black hole.  Since tidal interaction, which transfers energy and
angular momentum between spin and orbit, is taking place in a very
small region of space-time, it might not be unreasonable to expect
that a full relativistic treatment of the problem will not
qualitatively change this result, although it is clear that a number
of detailed questions still need closer attention. In particular,
Hut's analysis should be extended to a relativistic regime above resonance,
and the \citet{Gomboc05} analysis would have to include higher
order modes, mode splitting due to rotation, and a relativistic regime.
\section{Conclusions}
In this paper we have investigated the fate of small bodies, like
asteroids and comets, that may find themselves on highly eccentric
orbits in the vicinity of the Galactic black hole. Extending Hut's
analysis of the tidal evolution of close binary systems, we find that
they experience very high tides at their periastra, which heat and
eventually liquefy them. Those objects that are electrically
conductive are likely to have enough magnetic rigidity to
efficiently transfer angular momentum from orbit to spin, so that
the loss of orbital angular momentum leads to smooth transition from
bound to plunging orbits. We propose this as a mechanism that could
bring relatively small and condensed clumps of material to the
vicinity of the black hole. Such clumps can produce individual
accretion events characterized by the time scale of the last
circular orbit. The tidal energy released during this process can
reach up to $0.1\ mc^2$. This means that both the available energy
and the characteristic timescale are consistent with the energy
and timescale characteristic of Galactic flares.
\begin{acknowledgements}
We acknowledge support from the bilateral protocol of scientific and
technological cooperation between Italy and Slovenia.
 A.\ \v C.  acknowledges partial support from the grant
1554-501 of the Slovenian Research Foundation. A.\v C. and M. C.
acknowledge the hospitality of the Instituto de Astrofisica de Andalucia
(Granada, Spain), where part of this work was done. We thank the
anonymous referee for providing comments, which helped improve the
contents of this paper.
\end{acknowledgements}
\bibliographystyle{aa}

\begin{thebibliography}{}
%
\bibitem[Alexander(2005)]{2005PhR...419...65A} Alexander, T.\ 2005,
\physrep, 419, 65
\bibitem[Ashtekar \& Krishnan (2004)]{Ash}Ashtekar, A., Krishnan, B, 2004, Living Rev. Relativity 7, 10, http://www.livingreviews.org/lrr-2004-10
\bibitem[B{\'e}langer et al.(2006)]{2006JPhCS..54..420B} B{\'e}langer, G.,
Terrier, R., de Jager, O.~C., Goldwurm, A., \& Melia, F.\ 2006,
Journal of Physics Conference Series, 54, 420
\bibitem[Cochran et al.(1995)]{1995ApJ...455..342C} Cochran, A.~L.,
Levison, H.~F., Stern, S.~A., \& Duncan, M.~J.\ 1995, \apj, 455, 342
\bibitem[Dermott et al.(1988)]{1988Icar...76..295D} Dermott, S.~F.,
Malhotra, R., \& Murray, C.~D.\ 1988, Icarus, 76, 295
\bibitem[Eckart et al.(2006)]{2006A&A...455....1E} Eckart, A., Sch{\"o}del,
R., Meyer, L., Trippe, S., Ott, T., \& Genzel, R.\ 2006, \aap, 455,
1
\bibitem[Eisenhauer et al.(2005)]{Eisenh05} Eisenhauer, F.,
Genzel, R., Alexander, T., et al, 2005, \apj, 628, 246
\bibitem[Fang \& Lovelace(2005)]{2005PhRvD..72l4016F} Fang, H., \&
Lovelace, G.\ 2005, \prd, 72, 124016
\bibitem[Genzel et al.(2003)]{2003Natur.425..934G} Genzel, R., Sch{\"o}del,
R., Ott, T., Eckart, A., Alexander, T., Lacombe, F., Rouan, D., \&
Aschenbach, B.\ 2003, \nat, 425, 934
\bibitem[Goldwurm(2007)]{2007Goldwurm} Goldwurm, A. 2007, Comptes Rendus Physique, 8, 35
\bibitem[Gomboc \& \v Cade\v z(2005)]{Gomboc05} Gomboc, A., \& \v Cade\v
z, A. 2005, \apj, 625, 278
\bibitem[Hopman \& Alexander(2006)]{2006ApJ...645.1152H} Hopman, C., \&
Alexander, T.\ 2006, \apj, 645, 1152
\bibitem[Hut(1980)]{1980A&A....92..167H} Hut, P.\ 1980, \aap, 92, 167
\bibitem[Hut(1981)]{1981A&A....99..126H} Hut, P.\ 1981, \aap, 99, 126
\bibitem[Hut(1982)]{1982A&A...110...37H} Hut, P.\ 1982, \aap, 110, 37
\bibitem[Kivelson et al.(1993)]{1993Sci...261..331K} Kivelson, M.~G.,
Bargatze, L.~F., Khurana, K.~K., Southwood, D.~J., Walker, R.~J., \&
Coleman, P.~J.\ 1993, Science, 261, 331
\bibitem[Meyer et al.(2006)]{2006A&A...460...15M} Meyer, L., Eckart, A.,
Sch{\"o}del, R., Duschl, W.~J., Mu{\v z}i{\'c}, K., Dov{\v c}iak,
M., \& Karas, V.\ 2006, \aap, 460, 15
\bibitem[Misner, Thorne \& Wheeler (1971)]{MTW}  Misner, C.W.,  Thorne,
K.S., \&  Wheeler, J.A. 1973, Gravitation; Freeman and Company
\bibitem[Narayan(2000)]{2000ApJ...536..663N} Narayan, R.\ 2000, \apj, 536,
663
\bibitem[Paumard et al.(2006)]{2006ApJ...643.1011P} Paumard, T., et al.\
2006, \apj, 643, 1011
\bibitem[Xu et al.(2006)]{2006ApJ...640..319X} Xu, Y.-D., Narayan, R.,
Quataert, E., Yuan, F., \& Baganoff, F.~K.\ 2006, \apj, 640, 319
%
%
\end{thebibliography}
\end{document}